\documentclass[12pt,preprint]{aastex}
\usepackage{spr-astr-addons}

\newcommand {\kms}{km~s$^{-1}$}

\newcommand {\gsim}{\mbox{$\:\stackrel{>}{_{\sim}}\:$} }

\begin{document}

\title{Re-ionizing the Universe without Stars}
\shorttitle{Re-ionizing the Universe without Stars}
\shortauthors{Dopita \& Krauss}

\author{ Michael A. Dopita\altaffilmark{1}\altaffilmark{2}\altaffilmark{3},  Lawrence M. Krauss\altaffilmark{4}\altaffilmark{1} Ralph S. Sutherland\altaffilmark{1}, Chiaki Kobayashi\altaffilmark{1} \& Charles H. Lineweaver\altaffilmark{1}\altaffilmark{5}}
\email{Michael.Dopita@anu.edu.au}
\altaffiltext{1}{Research School of Astronomy and Astrophysics, Australian National University, Cotter Rd., Weston ACT 2611, Australia }
\altaffiltext{2}{Astronomy Department, King Abdulaziz University, P.O. Box 80203, Jeddah, Saudi Arabia}
\altaffiltext{3}{Institute for Astronomy, University of Hawaii, 2680 Woodlawn Drive, Honolulu, HI 96822, USA}
\altaffiltext{4}{School of Earth \& Space Exploration  \& Dept. of Physics, Arizona State University, Box 871504, Tempe, AZ 85287-1504, USA}
\altaffiltext{5}{Research School of Earth Sciences, Australian National University, ACT 0200, Australia}

\email{Michael.Dopita@anu.edu.au}
\begin{abstract}
Recent observations show that the measured rates of star formation in the early universe are insufficient to produce re-ionization, and therefore, another source of ionizing photons is required. In this \emph{Letter}, we examine the possibility that these can be supplied by the fast accretion shocks formed around the cores of the most massive haloes ($10.5< \log M/M_{\odot} < 12$) on spatial scales of order 1~kpc. We model the detailed physics of these fast accretion shocks, and apply these to a simple 1-D spherical hydrodynamic accretion model for baryonic infall in dark matter halos with an Einasto density distribution. The escape of UV photons from these halos is delayed by the time taken to reach the critical accretion shock velocity for escape of UV photons; 220 km~s$^{-1}$, and by the time it takes for these photons to ionize the surrounding baryonic matter in the accretion flow. Assuming that in the universe at large the baryonic matter tracks the dark matter, we can estimate the epoch of re-ionization in the case that accretion shocks act alone as the source of UV photons. We find that 50\% of the volume (and 5-8\% of the mass) of the universe can be ionized by $z \sim 7-8$.  The UV production rate has an uncertainty of a factor of about 5 due to uncertainties in the cosmological parameters controlling the development of large scale structure.  Because our mechanism is a steeply rising function of decreasing redshift, this uncertainty translates to a re-ionization redshift uncertainty of less than $\pm0.5$. We also find that, even without including the UV photon production of stars, re-ionization is essentially complete by $z \sim 5.8$. Thus, fast accretion shocks can provide an important additional source of ionizing photons in the early universe.
\end{abstract}


\section{Introduction}
An understanding of what re-ionizes the Universe at the end of the ``Dark Ages'' has so far eluded us. Thus far, two sources of ionizing photons have been suggested; stars and quasars \citep{Loeb01}. Both of these present difficulties. In the case of quasars, \citet{Mortlock11a} find very few QSOs at $z\sim6$, and only one is found above $z > 7$ (Mortlock, 2011b; see also note in proof). More importantly, if there were many fainter quasars producing sufficient UV to reionize the universe at $z=10$, then they would overproduce the present-day unresolved soft X-ray background \citep{Dijkstra04}. 

More hopeful is the mechanism of re-ionization by massive stars. It is certainly true that star formation persists back in time into the early universe. The Lyman break technique for detecting high redshift galaxies pioneered by \citet{Steidel96,Steidel99} has been successfully pursued out to redshifts of $z \sim 6$ (\emph{e.g.} \citet{ Bunker04,Yan04, Bouwens06,Yoshida06}). With the advent of the Wide Field Camera 3 (WFC3) of the \emph{Hubble Space Telescope} this technique has been pushed out to redshifts $z = 8-9$ \citep{Bunker10}. In other words, we are now directly able to measure star formation rates during the epoch of re-ionization.

Earlier it was thought that the epoch of re-ionization occurred at about $z_{\rm reion} \sim 6-8$. However, the more recent WMAP results \citep{Hinshaw09} place the re-ionization as early as $z_{\rm reion}=10.6 \pm 1.4$ \citep{Komatsu11}. This causes problems because the measured star formation density is found to decrease too quickly with increasing redshift. This problem is emphasized by  \citet{Bunker10}, who found that that the Lyman continuum photon density arising from star formation is in general insufficient tto reionize the Universe.  More recently, \citet{Bouwens11} found specifically that the ultraviolet flux available from galaxies at $z \approx 10$ is only $\sim 12^{+26}_{-10}$\% of what is needed for stars to be the reionizing source.

Given the lack of sufficient UV stellar ionizing radiation - even neglecting the issue of what fraction of this escapes from the concentrations of matter in which star formation is occurring, we need to consider whether there might be another source of ionization which has so far been neglected. An obvious candidate is to convert the mechanical energy of infall associated with the growth of structure in the early universe into UV photons. An early attempt to estimate UV photon production by this process was by \citet{Sasaki93}, but their physical model of the UV photon production was incomplete,and their mechanism relied upon (now ruled-out) isocurvature models, which provide much more small-scale structure than the concordance $\Lambda$CDM used today. 

\citet{Miniati04} considered the production of a thermal UV background radiation by turbulence and shocks associated with cosmic structure formation, and concluded that thermal photons alone are enough to produce and sustain He II reionization at $z\sim 6$. However, unlike the scenario considered here, the energy release was in the bulk of the baryonic gas rather than being concentrated in the highly unusual conditions that exist at the cores of the most massive agglomerations of mass in the early universe.

In this paper, we demonstrate that infall of baryonic matter into the dense cores of (very rare) massive haloes in the early universe creates fast, radiative accretion shocks. These generate ionizing radiation in the cooling zone which can ionize the accreting gas stream and escape into the inter-halo space beyond. The theory of such ``auto-ionizing'' fast shocks has previously been developed in the context of plasmas of normal chemical composition \citep{Dopita95,Dopita96} and applied to the study of emission regions associated with powerful radio jets and Seyfert galaxies \citep{Bicknell97,Allen98,Bicknell00}. Here, we show that fast, ``auto-ionizing'' accretion shocks are very efficient emitters of UV photons. We examine how well these shocks, acting alone, could ionize the early universe. 

We do not doubt that there may also exist a stellar UV continuum to assist the re-ionization, but here we will demonstrate that shocks produce a very significant number of additional photons at high redshift. In Section 2, we will discuss the radiative properties of these fast accretion shocks. In Section 3, we examine the circumstances under which such shocks form in the early Universe, and in Section 4, we examine the escape of these photons from the parent halo. In Section 5 we apply this theory to estimating the epoch of accretion shock driven re-ionization, and in Section 6 we compare this source of photons to the known star formation rate at $z\sim8$.

\section{Auto-Ionizing Fast Accretion Shocks}
The theory of fast shocks developed by \citet{Dopita95,Dopita96} and \citet{Allen08} cannot be applied to the accretion shocks formed in the early universe, because the infalling gas is nearly pristine, with a very low abundance of heavy elements. Therefore we have run a series of fast shock models using the MAPPINGS 3s code, an updated version of the code which was described in \citet{Sutherland93}. We have varied the shock velocity, $v$, between 100 and 500 \kms. We assume the shock to be running into a medium which is in collisional ionization equilibrium at 30000~K.  The abundances adopted are one thousandth of solar where the solar abundances are taken from \citet{Grevesse10}.  These models are assumed to be free of magnetic fields.

\begin{figure}[h]
\includegraphics[width=0.9\hsize]{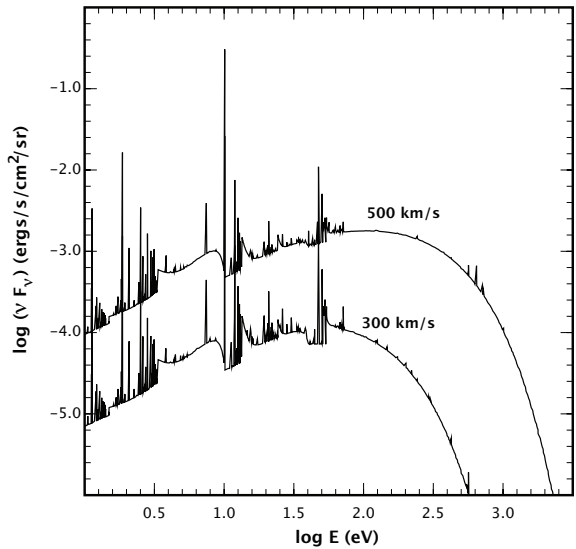}
\caption{The optical, UV and soft X-ray spectrum of fast shocks with velocities of 300 \kms and 500 \kms. The spectra show the resonance lines of hydrogen and helium, and the 2-photon and bound-free continua from these species, superimposed upon a thermal electron Brehmsstrahlung continuum. Such shocks might be detectable in their Lyman-$\alpha$ and confirmed by the presence of HeII 1640\AA\ emission.}\label{fig_1}
\end{figure}

The optical and UV spectrum of two of these models is shown in Figure \ref{fig_1}. The spectra are dominated by the hydrogen and helium (He I and He II) resonance and recombination lines, with bound-free and two-photon continua superimposed on an underlying electron brehmsstrahlung spectrum. These shock spectra are softer than those computed for plasmas with more normal abundance because the channels for cooling the plasma through the collisonal excitation of heavy element resonance lines and continua are absent. In particular, these shocks produce many He II photons which efficiently ionize H I and He I. The relatively `soft' UV spectrum produced in a fast shock with primordial composition couples better to the opacity of the un-ionized gas, and is therefore more effective in producing ionization in the surrounding medium.

The cooling timescale of the hot plasma in the shock velocity range 200-400 kms$^{-1}$ is $\sim 4\times 10^5/n$ yr, where $n$ is the hydrogen density in the pre-shock material. For the accretion flows we are considering, the density at the shock is estimated to be in the range $1 < n < 10$ cm$^{-3}$, the cooling timescale is short compared with the accretion timescale ($\sim 5\times 10^7$ yr). Furthermore the cooling length, $\sim 20/n$ pc, is small compared to the physical scale of either the accretion flow, or the turbulent disk of shocked gas formed. We can therefore treat both the shock and its UV photon emitting zone as a thin shell.

In a plane-parallel shock, 50\% of the UV photons go to generate a photoionized tail in the cooling zone, while the other 50\% escape upstream into the accretion flow.  The number of upstream ionizing photons $\phi(v)_{\rm up}$  produced per hydrogen atom processed through the shock  can be adequately represented from our radiative shock models as:
\begin{eqnarray}
\phi(v)_{\rm up} = 0.2v_{100}^2; ~100<v<280 {\rm km/s}  \nonumber \\
\phi(v)_{\rm up} =10v_{400}^5; ~ 280<v<400  {\rm km/s}  \label{phi}\\
\phi(v)_{\rm up} =10v_{400}^2; ~ v> 400 {\rm km/s}  \nonumber 
\end{eqnarray}
where $v_{100}$ and $v_{400}$ are the shock velocities normalized to 100 and 400 \kms , respectively.

While $\phi(v)_{\rm up} < 1$, the upstream photons are absorbed in a thin photoionized precursor which is stationary with respect to the shock. When $\phi(v)_{\rm up} > 1$, the velocity of the ionization front in the accretion flow exceeds the shock velocity and the ionization front is free to propagate upstream. Since the density of the accretion flow falls with radius from the accretion shock, the ionization front will accelerate with time, rapidly ionizing the accretion flow and escaping into the low density medium. Our accretion shock mechanism for re-ionization has the attractive feature that the source of the ionization (at the accretion shock) is located on the surface of the largest baryonic mass concentrations in the early Universe. The escape of these ionizing photons is therefore much more efficient than for ionizing photons produced by deeply embedded regions of star formation. From Equation \ref{phi} the critical shock velocity for pre-ionization of the accretion flow is  $v \approx 220$\kms. This is appreciably larger than the corresponding condition for plasmas of solar composition ($\approx 175$ \kms), because the heavy element line and continuum cooling of highly ionized species more readily produce ionizing photons in the hotter zones of the shock than does gas of primordial composition. These fast shocks produce a radiation field with much higher ionization parameter in the precursor gas than would stars. The ionization parameters, $0.001 < U < 0.02$, are comparable with those produced by AGN, although their soft X-ray fluxes are much less.
 
 \section{Making UV photons in massive halos}
 In our model, the fast accretion shocks needed for re-ionization can only be formed around the boundaries of the most massive collapsed agglomerations of baryonic matter in the early Universe. These mass concentrations must in turn be located at the cores of the most massive halos that can exist at a given epoch. Given that there is little star formation, there is also little or no feedback to prevent the collection of baryonic matter in the core. This (nearly radial) infall of baryonic gas is also found in sophisticated simulations such as those of the OWLS project \citep{Sales10}. 
 
 In the standard cold accretion model, the  flow remains cool as long as the compressional heating rate and the heating rate by turbulence is matched by the cooling rate, locally. In a large fraction of the Universe such cold accretion flows have their inflow velocity limited by the compressional heating and the need to radiate this heating away. However, in massive halos, pressure balance is lost when the cooling timescale becomes shorter than the accretion timescale. In `cold'  flows the gas is near $10^4$K and the cooling timescale is of order the recombination timescale, $\tau_{\rm rec} \sim 10^5/n$ yr, where $n$ is the hydrogen density (cm$^{-3}$). Since the free-fall timescale is of order $10^8$ yr, the gas loses hydrostatic support and free-falls wherever the density exceeds$\sim 10^{-3}$ cm$^{-3}$. In the halos we are to consider, this occurs at about 6 scale radii or ~50 kpc. Within this radius, the gas is effectively pressure-free and in free-fall, achieving velocities of several hundred km s$^{-1}$ by the time it reaches the halo core.
 
In this picture of  rapid accretion onto a dense baryonic core in the dark matter halo, the production of ionizing photons,  $\dot{N}$, is  given by:
 \begin{equation}
 \dot{N} = \phi(v)_{\rm up}\frac{\dot{\cal M}} {\mu m_{\rm H}} ~~{\rm s^{-1}} \label{photons}
 \end{equation}
 where $\dot{\cal M}$ is the mass accretion rate, $\mu$ is the mean atomic weight of the infalling gas, $m_{\rm H}$ is the mass of a hydrogen atom.  For each halo, we therefore need to compute both $\dot{\cal M}$ and the velocity of accretion onto the baryonic core, which by use of equation \ref{phi} will then give us $ \phi(v)_{\rm up}$.
 
For the halos, we use the Einsato density profile \citep{Navarro04,Springel08} for the DM density, in terms of a scaled radius $x = r/r_s$:
\begin{equation}
\rho (x) = \rho_{-2} \exp\left[ \frac{-2}{\alpha} (x^\alpha - 1) \right] \ .
\end{equation}
In this equation, the logarithmic derivative is taken to be equal to $-2$ at the scale radius, so that $\rho(1) = \rho_{-2}$.
The parameter $\alpha$ has a small variation about $\alpha = 0.18$, which is the value adopted here, and $\rho_{-2}$ is chosen to give the halo mean density inside the virial radius, $\rho_{\rm halo}$, once the physical virial radius and halo over--density are known.

The overall scaling is obtained by requiring $\rho_{\rm halo}$ to be related to the critical density of the Universe at the redshift considered,
\begin{equation}
	\rho_c = \frac{3 H_0^2}{8 \pi G} \left[{\Omega_\Lambda + \Omega_M (1+z)^3}\right] \, .
\end{equation}
Assuming the collapse of a spherical top hat perturbation, the ratio of the mean halo density to the critical density of the Universe, 
$\Delta_c = \rho_{\rm halo}/\rho_{c}$, is given by \citep{Bryan98}; 
\begin{equation}
	\Delta_c = 18\pi^2 + 82f(z) - 39f(z)^2\, ,
\end{equation}
where
\begin{equation}
f(z) = \left[ \frac{\Omega_M (1+z)^3}{\Omega_\Lambda + \Omega_M (1+z)^3} \right]-1\, .
\end{equation}
We take $ \Omega_M = 0.2715$ and
$ \Omega_\Lambda = 0.728$ from the WMAP+BAO+$H_0$ maximum likelihood value in the WMAP seven year analysis \citep{Komatsu11}.

The value of the virial radius $x_{\rm vir}$ in units of the scale radius is obtained using a fit by \citet{Duffy08}, (termed as concentration in that work) with the desired virial mass for the appropriate redshift $z$, and cosmology (here $h = 0.704$) :
\begin{equation}
x_{\rm vir} = 8.82 (1+z)^{-0.87}   \left[ \frac{M_{\rm vir}}{2.0\times 10^{12} h^{-1}}\right]^{-0.106} \, .
\end{equation}
Once the virial radius is known in terms of the scale radius, the physical scale is chosen so that the mass of the halo averaged over the virial volume gives the desired mean halo density. The gravitational potential is then calculated integrating  $\Phi(r) = GM(r)/r$.  An analytical form for this potential can be readily derived.

As an approximation of the complex physics involved in the collapse process, we assume that the baryonic matter follows the dark matter until virialization of the dark matter halo, at which point the baryonic matter begins to fall freely towards the centre of the halo. Thus, in our initial configuration, the baryonic mass starts off from rest and with the  same radial density profile as the dark matter halo. The baryonic matter is therefore already bound within the potential of the dark matter halo, and therefore any kinetic energy of infall to the initial radius during the assembly of the halo is effective assumed to have been already dissipated - presumably through the low velocity shocks associated with the halo assembly in the cold accretion scenario.  

From this (somewhat uncertain) approximation to the initial state, we have followed the subsequent accretion in 1-D spherical coordinates using the Lagrangian hydrodynamic code, \emph{Fyris Alpha} \citep{Sutherland10}. This code includes all relevant gas physics and gravitational fields. By using 5 nested domains, a final resolution of 0.3 pc is achieved in this simulation. This should be compared with the $\sim 0.5$ kpc spatial resolution achieved in the best smoothed particle hydrodynamics (SPH) cosmological codes.

As expected, within the critical radius at which pressure support is lost due to fast cooling (within about $R \sim 50$ kpc as estimated above) the baryonic matter effectively free-falls towards the centre, since it has no thermal pressure support, and very little rotation. In the vicinity of the nucleus, the infalling gas is shocked, and collects into a small, dense, rotationally-supported disk whose size is determined by the torquing of the parent halo. The vertical scale height of this disk is determined by thermal and turbulent support. Its pressure is given by the ram pressure of the accretion flow, since the radiative shocks can be treated as isothermal. Here we assume that the disk has a rotational velocity similar to the free-fall velocity in the dark matter halo. Thus the depth of the central potential (baryonic plus dark matter) is approximately doubled, and the velocity associated with the accretion shock is $\approx \sqrt2 \times$ the free-fall velocity in the dark matter halo.

\begin{figure}[h]
\includegraphics[width=0.9\hsize]{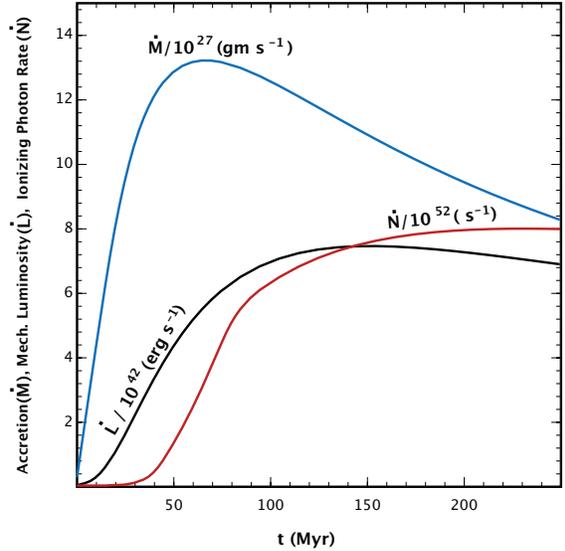}
\caption{For a halo with mass $10^{11}$M$_{\odot}$ within the virial radius, we show the mass accretion rate, mechanical luminosity through the accretion shock, and the rate of production of ionizing photons as a function of time after the halo first forms. Note (at late times) how a decreasing mass accretion rate,  $\dot{\cal M}$, is compensated by an increase in the accretion velocity so that the rate of production of ionizing photons remains essentially invariant with time. However, the number of ionizing photons produced per baryon accreted is only of order 0.1\% of what would be produced if these baryons were all formed into stars.}\label{fig_2}
\end{figure}

For each (non-rotating) spherical halo, the time dependence of the accreted mass, the free-fall accretion rates, and the free-fall velocity at $r\sim 1.0$kpc  were calculated. Using equation \ref{photons} we computed the rate of production of ionizing photons launched into the accretion flow. In figure \ref{fig_2} we show a representative halo with mass $10^{11}$M$_{\odot}$ within the virial radius. The peak mass accretion rate is reached at 50-60 Myr, and corresponds to $205$M$_{\odot}$yr$^{-1}$, when the accretion velocity is $320-350$km~s$^{-1}$. Because we do not properly model the initial formation of the halo, it is not clear whether photons start to escape the shock contemporaneously with this formation, or whether we have to wait an additional $\sim 50$Myr as indicated in figure \ref{fig_2}. We therefore compute both cases.

The rate of production of ionized photons continues to increase with time, as the accretion velocities continue to rise. For more massive halos   $\dot{\cal M} \propto M$, and since $v \propto M^{1/2}$,  equation \ref{photons} along with equation \ref{phi} shows us that $\dot{N} \propto M^2$. For less massive halos, the rate of production of UV photons falls steeply, due to both the steeper dependence of $\dot{N}$ on velocity, and the fact that $v(t) < 220$~km s$^{-1}$ for an extended period of time from the start of the collapse.

 \section{The escape of photons from the halos}
It is not enough for the accretion shocks to produce an excess of UV photons to provide re-ionization. The photons have to first ionize the gas in the halo, and even when this has been done, we still have to account for the number of photons being used per unit time in maintaining the ionization of the halo. In our numerical models, we find that, over a wide range in radius, the number density $n(r)$ of  infalling gas in our Einsato density profile dark matter halos can be approximated by a power law $n(r) \propto r^{-1.83}$. Since the number of recombinations per unit volume is $\propto n(r)^2$, the number of photons needed to maintain ionization in a spherical shell of unit thickness in the accretion flow falls off $\propto r^{-1.66}$. Thus, the photons required to maintain ionization rapidly become unimportant, and once the ionization front moves away from the vicinity of the accretion shock, it rapidly accelerates outwards. 

The ionizing photons must still ionize all of the baryonic matter in the halo before they can escape into the universe at large. The number of photons required is considerable. For a halo with mass $10^{11}$M$_{\odot}$ within the virial radius, the total baryonic mass to infinity is $1.4 \times 10^{11}$M$_{\odot}$. We compute that, once the UV photons start escaping from the vicinity of the shocks ($\sim 50$ Myr after the formation of the halo), it takes a further  $\sim 60$ Myr to complete the ionization of the baryonic halo. For more massive halos, $\dot{N} \propto M^2$ so that  the timescale taken to ionize the halo falls  $\propto M^{-1}$. The steep mass dependence of the photon production coupled with the decreasing ionization latency time with mass allows the more massive halos to successfully compete for the net UV photon production, despite their much lower space densities.

Detailed hydrodynamic modeling of the baryonic gas \citep{Greif08, Goerdt10} shows that it is accreted in streams, rather than isotropically. This is the so-called ``cold accretion'' model. In this, the early escape of photons is favored. This is because the specific photon flux scales as the mass accretion rate per unit area, so more photons per unit area are produced in denser accretion streams. These photons can then easily escape between these accretion streams, resulting in more prompt re-ionization.

 \begin{figure}[h]
\includegraphics[width=0.9\hsize]{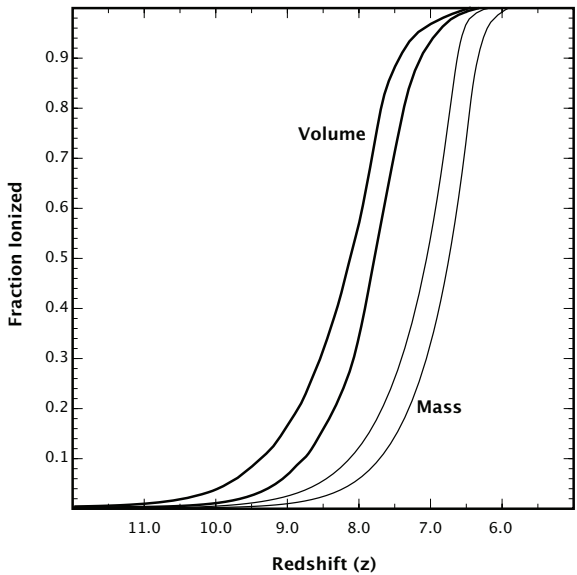}
\caption{The fraction of the volume, and of the total mass, of the universe that is ionized by accretion shock-produced UV photons as a function of redshift,. We use the new high-resolution cosmological calculation of Kobayashi, (see \citet{Kobayashi07} for the description of an earlier version of this code). The ionization model used is a photon diffusion approximation in which the ionization fronts are assumed to move rapidly to ionize the least-dense phases first. The number of ionizing photons needed to maintain the ionization in the ionized volume are explicitly included in this calculation. We have computed two cases - that in which the shocks photon production is contemporaneous with the formation of the halo (left-hand curves) and the case in which this photon production is delayed by 50 Myr (right-hand curve). According to our model,  50\% of the volume (5-8\% by mass) of the universe is ionized by $z \sim 8$, and ionization is essentially complete by $z=5.8$.}\label{fig_3}
\end{figure}

\section{Reionizing  the Universe}  
The rate of production of ionizing photons be co-moving  Mpc$^{3}$ can be calculated using the density of halos as a function of mass and redshift given by, for example, the simulations of \citet{Mo02} or \citet{Jenkins01}. Equation \ref{photons} gives the photon production rate for a halo of a given mass. We have computed two cases - that in which the UV photon production in shocks is contemporaneous with the formation of the halo and the case in which this photon production is delayed by 50 Myr, related to the free-fall collapse time. 

The photon production rate is summed over the space density of halos given by \citet{Mo02} to give the production rate of photons per co-moving Mpc$^{3}$. We then compute the time dependent ionization of the universe using a photon diffusion approximation. In this, the ionization fronts move rapidly to ionize the least-dense phases first, and the denser HI in the halos is ionized more slowly and at later times. This provides a reasonable approximation to this complex process \citep{Iliev09}. The photons needed to maintain the ionization in the ionized volume are included in the calculation. The appropriate densities have been computed for us using a more recent (higher resolution) version of the cosmological simulation by  \citet{Kobayashi07}. The result is shown in Figure \ref{fig_3}. We find that with the \citet{Mo02} cosmological parameters, $\sim 50$\% by volume (and 5-8\% by mass) of the universe is ionized by $z \lesssim 8$, and that re-ionization is complete by  $z\gsim 5.8$. Note that the effect of more prompt escape of UV photons would be to markedly increase the degree of ionization at early times. At a redshift of $z =10$ the photon density is increased by more than a factor of three by this prompt escape.

Note that the space density of massive halos given by the equations of \citet{Jenkins01} and the models of \citet{Mo02} assume the cosmological parameter $\sigma_8$. This is related to the mean mass density fluctuations on scales of $8h^{-1}$ Mpc today, and is $\approx 0.9$.  This enters exponentially into the overall collapse fraction.  There is some uncertainty in this value and the more recent estimates \citet{Komatsu11} suggest $\sigma_8 \approx 0.8$.  If one assumes this latter value, the fraction of baryons in massive halos at high redshift will be reduced by approximately a factor of 4.5 compared to those estimated using \citet{Mo02}.  As a consequence, the rate of production of ionizing photons will be reduced by a similar factor. However, the effect on the epoch of re-ionization is much smaller, as we now show.

As described above, only the most massive halos are important as sources of ionizing photons. We estimate that  dominant halos for production of UV photon s have masses in the range $10^{11-11.5}$ M$_\odot$. The photon production by accretion shocks increases very rapidly with decreasing redshift, thanks to the rapid increase in space density of halos, and the increasing importance of the more massive halos. At $z \sim 8$, it is increasing  (with decreasing redshift) $\propto (1+z)^{-10}$.  At the same time the number of photons needed to ionize the universe decreases $\propto (1+z)^{3}$. These factors serve to ``lock in'' the epoch of re-ionization into a rather narrow range in $z$. Furthermore, as a result of this strong redshift dependence, changing from the \citet{Mo02} cosmological parameters to the ones favored by \citet{Komatsu11} would reduce the onset of ionization in redshift by only $\delta z \approx 0.5$.

\section{Comparison with star formation rates}
The accretion shocks provide $7 \times 10^{52}$  ionizing photons s$^{-1}$ for a halo mass of $\log M/M_{\odot} = 11$, and as many as $8 \times 10^{54}$ ionizing photons s$^{-1}$ for a halo mass of $\log M/M_{\odot} = 12.0$. These fluxes are equivalent to the ionizing photons produced by 1400 and 160,000 O5 stars, respectively. These numbers may seem high, but correspond to the relatively modest total star formation rates of 0.07 $M_{\odot}$~yr$^{-1}$ and 8 $M_{\odot}$~yr$^{-1}$, respectively. Allowing for the space densities of the halos of the appropriate mass, at a redshift of $z=8$ the UV photon density provided by a  halo mass of $\log M/M_{\odot} = 11$ is equivalent to a star formation density of $10^{-3.75}M_{\odot}$~yr$^{-1}$~Mpc$^{-3}$. For a halo mass of $\log M/M_{\odot} = 11.5$, this rises to $10^{-3.2}M_{\odot}$~yr$^{-1}$~Mpc$^{-3}$ and by $\log M/M_{\odot} = 12$, has fallen to $10^{-4.1}M_{\odot}$~yr$^{-1}$~Mpc$^{-3}$. The total effective equivalent star formation density of shocks at $z=7-8$ is  $(1-2) \times 10^{-3}M_{\odot}$~yr$^{-1}$~Mpc$^{-3}$, between half and all of the currently measured star formation density at this epoch \citep{Bunker10,Bouwens11}. Thus, even if accretion shocks do not dominate the UV photon budget in the early universe, they remain an important contributor to the total.

\section{Conclusions}
We have shown that re-ionization can, in principle,  be driven by the UV photons produced by fast, radiative and ``auto-ionizing'' accretion shocks associated with the most massive haloes formed at any epoch in the early Universe. These shocks are located at the outer edges of the most massive agglomerations of baryonic matter, and therefore the UV radiation produced in such shocks can readily escape into inter-galactic space. According to this model, the minimum mass halo that can produce an appreciable UV flux has a mass $\gsim 3 \times 10^{10}M_{\odot}$. Such halos are not formed at appreciable space density until  $z\sim 12$, so the epoch of re-ionization cannot start before this.  The production of ionizing photons increases very rapidly thereafter, while the number of photons required to complete re-ionization falls. Thus, this process provides a epoch of re-ionization which is rather sharply defined. We find that 50\% of the volume (and 5-8\% by mass) of the universe is ionized by $z \lesssim 7-8$, and that re-ionization is complete by  $z\gsim 5.8$. The UV photon density produced by these accretion shocks at $z\sim 7-8$ is approximately half that produced by the observed star formation at this redshift, so even if accretion shocks do not dominate, they remain an important contributor to the total UV photon production budget. In addition, the accretion shocks are formed around the periphery of the main mass agglomeration, and so the escape of the UV photons is favored, especially when the gas accretion is in cold streams. 

The stars, on the other hand, will be predominantly formed within these dense regions where the pressure is of order $P/k \sim 3\times 10^8$~cm$^3$K. Under these conditions, individual the H~II regions around O or B stars are only  $\sim 0.5$~pc in diameter \citep{Dopita06}. Individually these stars will be ineffective in ionizing the universe. Collectively, they could blow out of the dense disk gas, and UV photons could then escape. However, the timescale for this to occur is appreciably longer than the baryon free-fall time and is unlikely to impact upon our estimate for the epoch of re-ionization.  A doubling of the UV radiation density to include a possible stellar contribution would push our estimate of the epoch of re-ionization redshift back by $z \sim 0.6$ in redshift.

To resolve the accretion shocks will take much higher resolution than has hitherto been achieved in simulations. The thickness of the hot shocked gas is only $\sim 10$pc. Current cosmological models have typical particle sizes of $10^6$M${\odot}$ and a maximum spatial resolution $\sim 500$pc. Since the infalling gas has a cooling timescale much shorter than the infall timescale, it is not hotter than $10^4$K. The post-shock turbulent gas has a similar temperature, and so current simulations will make it appear that the gas is accreted entirely in ``cold mode''.

More precise estimates of the epoch of re-ionization will have to await numerical hydrodynamic simulations which correctly compute the temporal and mass dependence of the baryonic fast accretion shocks, which explicitly compute the propagation of the ionization fronts, which compute the UV photon production rate from individual halos, and which correctly model the growth of the nuclear disk including disk rotation and turbulence. Such models are in preparation, and these models will serve to determine to what extent these proto-galactic shocks are observable, either through their Lyman-$\alpha$ or their HeII 1640\AA\ emission lines or through the soft thermal X-ray continuum which they produce.

\begin{acknowledgments}
We thank both the referee and Sir Martin Rees for their constructive comments, which have much improved this paper. Mike Dopita acknowledges the support of the Australian Research Council (ARC) through Discovery  projects DP0984657.  He would also like to thank Andrew Bunker for his inspiring seminar on star formation in the early universe, which triggered the ideas presented here. Krauss acknowledges Australian National University for its support and hospitality as a Distinguished Visiting Professor, as well as travel funds from Brian Schmidt, making possible this collaboration.   He also acknowledges support from the U.S. Department of Energy. Finally, we also acknowledge  Ben Moore for constructive discussions and his valued input.
\end{acknowledgments}
\bigskip \newline {\bf Note Added in proof:}

The discovery of a massive quasar at a redshift of $z=7.085$ by Mortlock et al. (2011b) has a number
of important implications to both this paper, and to existing comological models. First, as stated
by the authors, it establishes that there is an appreciable un-ionized fraction at this redshift, 
in contrast with the WMAP result (Komatsu et al. 2011). Second, its spectrum establishes that
the broad line region has chemical abundances similar to solar, even though the universe is only
0.77 Gyr old at this epoch, implying a large conversion of baryons into heavy elements in its vicinity. 
Third, the estimated black hole mass, $2 \times 10^9$M$_{\odot}$ implies very prompt collapse. If 
built up at the Eddington rate from an initial mass of order $100$M$_{\odot}$, we would need about 
$3 \times 10^8$ yr to do this! This implies that the host halo, $M \sim 10^{11}$M$_{\odot}$ was not 
only collapsed but had formed a super-dense core by $z \sim 10$, supporting the possibility of
monolithic collapse of massive halos in the early universe. Monolithic collapse, the production of 
massive black holes, as well as possibly concentrated regions of star formation suggests that all 
three sources of UV photons may be important for re-ionization, and that the current paradigm of 
massive halo formation is likely incomplete

\end{document}